\documentclass[12pt]{article}

\usepackage{PRIMEarxiv}
\usepackage{amsmath}
\usepackage[utf8]{inputenc} 
\usepackage[T1]{fontenc}    
\usepackage{hyperref}       
\usepackage{url}            
\usepackage{booktabs}       
\usepackage{amsfonts}       
\usepackage{nicefrac}       
\usepackage{microtype}      
\usepackage{lipsum}
\usepackage{graphicx}
\usepackage{wrapfig}
\graphicspath{{media/}}     
\usepackage{authblk} 
\usepackage{caption}
\usepackage{fancyhdr}
\usepackage{float} 
\usepackage[section]{placeins} 
\usepackage{placeins}

\begin{document}
\begin{center}
    {\LARGE \bf Role of spatial embedding and planarity in shaping the topology of the Street Networks} \\
    \vspace{1.8em}
    Ritish Khetarpal\textsuperscript{1}, Aradhana Singh\textsuperscript{1*} \\
    \textsuperscript{1} Department of Physics, Indian Institute of Science Education and Research Tirupati, India \\
    \texttt{ritishkheterpal12@gmail.com}, \texttt{aradhana22singh@gmail.com*} \\
    \vspace{1em}
    \today
\end{center}

\begin{abstract}
The topology of city street networks (SNs) is constrained by spatial embedding, requiring non-crossing links and preventing random node placement or overlap. Here, we analyzed SNs of $33$ Indian cities to explore how the spatial embedding and the planarity jointly shape their topology. Overall, we found that all the studied SNs have small-world properties with higher clustering and efficiency. The efficiency of the empirical networks is even higher than that of the corresponding degree of preserved random networks. This increased efficiency can be explained by Dijkstra's path-length distribution, which closely fits a right-skewed normal or log-normal distribution.   
Moreover, we observed that the connectivity of the streets is length-dependent: the smaller streets connect preferably to the smaller streets, while longer streets tend to connect with the longer counterparts. This length-dependent connectivity is more profound in the empirical SNs than in the corresponding degree preserved random and random planar networks. However,  planar networks maintaining the empirical spatial coordinates replicate the connectivity behavior of empirical SNs, highlighting the influence of spatial embedding. Moreover, the robustness of the cities in terms of resilience to random errors and targeted attacks is independent of the SN's size,  indicating other factors, such as geographical constraints, substantially influence network stability. 
\end{abstract}

\section*{Keywords}
Street Networks, Spatial small-world, Normal Distribution, Rich Club, Robustness 

\section{Introduction}

\begin{figure}[htp!]
    \centering
    \includegraphics[width=0.76\textwidth]{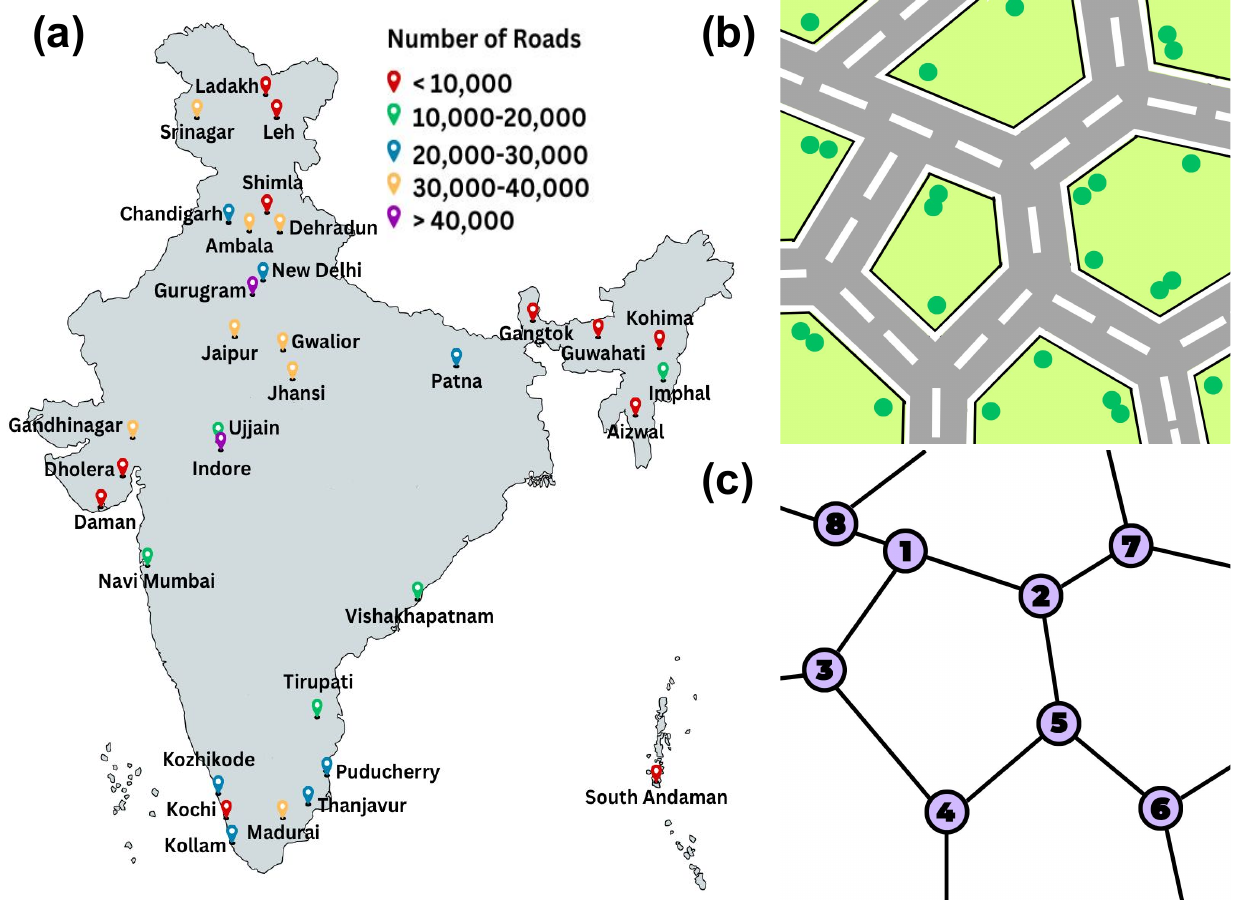}
    \caption{{\bf Street Network:} Sub-fig(a) displays the locations of the studied cities on the map of India. (b) A simplified street layout showing roads and intersections. (c) The corresponding network representation, where each intersection is modeled as a node, and the roads between them are modeled as links.}
    \label{fig:Street_network_visualization}
\end{figure}

\noindent 
The traffic dynamics of any urban area are influenced by the topology of its SN, and understanding its organization can help come up with better urban planning. The complexity of the SNs' comes from their vastness and spatiality,  making network theory a powerful tool for their analysis. The roads typically enclose blocks, and intersections appear visually distinct (Fig.~\ref{fig:Street_network_visualization}(b)). These intersections can be represented abstractly as nodes and the roads connecting them as edges, illustrating how a street map can be translated into a network structure 
( Fig.~\ref{fig:Street_network_visualization}(c)).
Previous studies on SNs have reported that their topology combines the tree-like and the lattice-like structures \cite{tree_like}. The tree structure represents a network without loops and is associated with better connectivity. The lattice topology features many loops and is associated with greater pedestrian safety and lower cost. Several factors influence the topology of the SNs, and urban planning is one of them. The planned cities have more grid-like behavior, resulting in the lower orientation entropy \cite{orientation_order_100_city}. Additionally, the topology may also change with time  \cite{entropy_40_cities, meshedness_paper}. For example, the evolution of the city of Dundee over more than $400$ years has shown a change in the length distribution, with the increase of length and the orientation entropy \cite{entropy_40_cities}.   
The topology of the SNs also varies significantly based on a country's geographical location and economy \cite{road_network_evolution_US_1900, land_use_chinese_cities, beijing_street_network}. Temporal changes in the topology are also associated with geographic factors. For instance, the evolution of the SNs of the United States over $115$ years has revealed a non-uniform influence by geography \cite{road_network_evolution_US_1900}. 

\noindent The SNs are generally planar, limiting the edges from crossing each other \cite{planarity_street_network,planarity_densification_exploration,planar_growing_random_london_network,planar_european_cities}. The planarity of the SNs has been shown to shape the topology, impose the constraints on their betweenness centrality, and make its distribution invariant of the topology and the spatial layout, a quality not shown by the non-planar networks  \cite{from_bc_street_network}. Additionally, the SNs come with another constraint: junctions can not be placed randomly. The current literature lacks the combined effect of planarity and spatial embedding on the topology of SNs. We study the SNs of $33$ cities across India to address this gap. We compare empirical SNs with the corresponding configuration networks, random planar networks, planar networks generated to preserve the spatial embedding, and the model spatial small-world networks. We find that the efficiency and clustering of the studied SNs are higher than the corresponding random configuration networks. Moreover, the associated costs for the SNs are less because the small streets are more abundant than the longer ones. Therefore, the studied SNs have the spatial small-world feature. Additionally, the distribution of the Dijkstra's for all the studied SNs follows the right skewed normal distribution or log-normal distribution. This finding accounts for the enhanced efficiency of the empirical SNs compared to their corresponding random configuration and the random planar networks. 
\noindent Furthermore, we observed a distinct pattern in how streets connect based on their length: the shorter streets tend to connect predominately with other short streets. In comparison, longer streets are prominently linked with similar or other long streets. This connectivity pattern is absent in random configuration, random planar, and model networks. However, it does appear in the planar networks generated using the geometric embedding of the corresponding empirical networks. These results suggest that both the geometric embedding and planarity significantly shape the topology of the SNs.
We also investigate the robustness of the SNs in terms of tolerance for errors and attacks. We find that the target removal based on the edge betweenness centrality similarly impacts the stability of all the studied SNs, regardless of their differing network size. However, errors have a more significant impact on the cities with higher altitudes. Indicating that the stability of SNs is influenced not only by their topology but also significantly by geographical factors. In the following, we discuss our results in detail.

\section{Results}
\subsection{Data Collection and Network construction}
\noindent We extract the street network data using one of the most popular geographic databases, OpenStreetMap (OSM)\cite{OpenStreetMap_reference}. The graphs generated from OSM represent unweighted, undirected networks where junctions act as nodes and streets serve as edges. Next, we obtained the coordinates (latitude and longitude) of the junctions and calculated the geodesic distance between them. We then assigned weights to the edges using the Python modules OSMnx\cite{OSMnx_main_paper} and NetworkX\cite{NetworkX_module}. 

\subsection{Small-worldness of SNs}
\begin{figure}[htp!]
    \centering
    \includegraphics[width=0.90\textwidth]{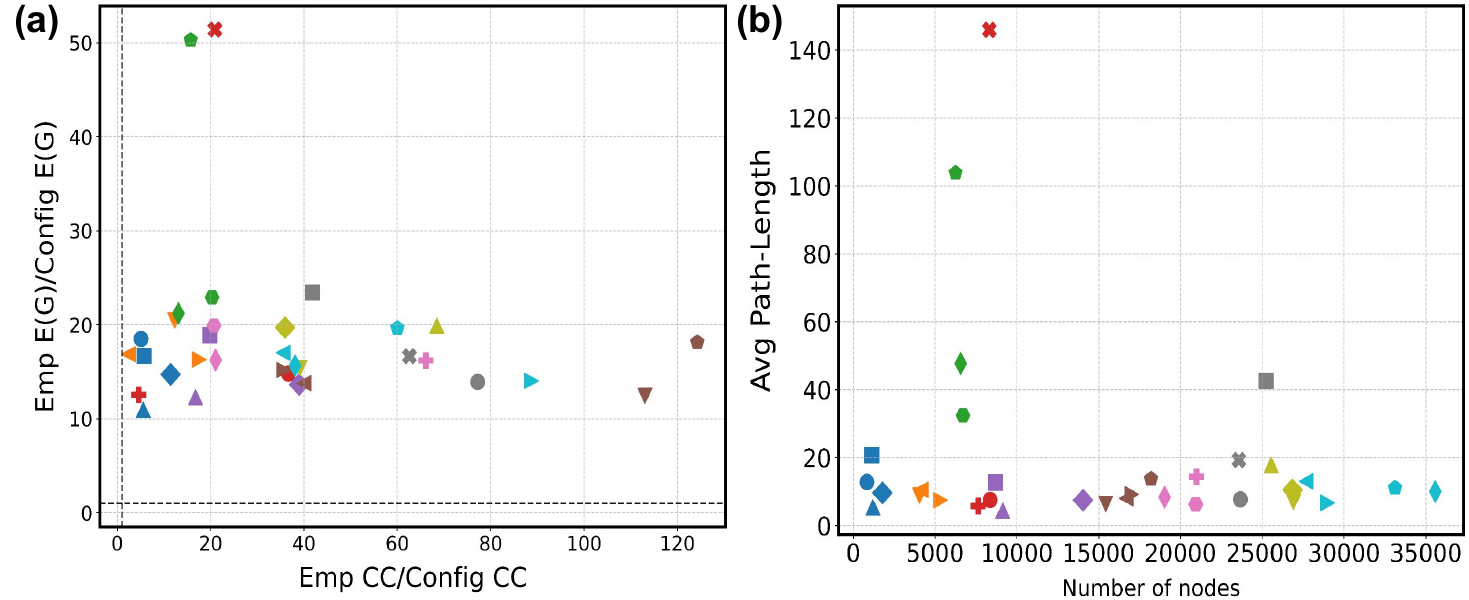}
    \caption{{\bf Small-world SNs:} Sub-fig(a) depicts normalized clustering versus the efficiency of empirical networks by the corresponding degree preserved configuration networks. Sub-fig(b) indicates no correlation between the network size and the average path length, indicating it depends on the network's topology.}
    \label{fig:Small_world_behavior}
\end{figure}

\noindent The small-world networks are characterized by high clustering, a trait shown by the regular networks, and high efficiency, which is the property of the random networks \cite{small_worldness_clustering_efficiency_ratio, strogats_small_world}. The SNs we study are sparse, so we compare their clustering and efficiency values with the corresponding degree of preserved random networks\cite{configuration_model_paper,random_model_original_paper}. 
To illustrate this comparison, we plot the clustering and efficiency values ratio in Fig.~\ref{fig:Small_world_behavior}(a). We find that the clustering and efficiency in the empirical network are more significant than that of corresponding degree-preserved configuration networks. This observation offers strong evidence of small-world behavior within the street networks, characterized by high clustering and efficiency \cite{small_world_k_clustering_sweden_street,dual_graph_main_paper}. Moreover, the average Dijkstra's path length is independent of the size of the cities and mainly depends on their topology (Fig.~\ref{fig:Small_world_behavior}(b)). We will discuss this in detail again in the coming sections. 
\noindent Next, we studied the Meshedness coefficient (Method) to confirm the existence of the grid-like structures within the SNs. A non-zero value of Meshedness for all the studied SNs indicates that all cities exhibit some degree of grid-like structure, suggesting a semi-lattice pattern in their layouts. However, the Meshedness coefficient is generally low across all cities, particularly for those in hilly regions (Table:~\ref{tab:network_data}). Additionally, planned cities like Chandigarh and Navi Mumbai exhibit higher meshedness coefficients than organic cities like Patna and Dehradun, indicating that organic cities have relatively more tree-like structures than planned ones. 
The average Meshedness for the studied cities is $0.1359$, closer to the previously reported average Meshedness for the SN of London than for the New York City \cite{meshedness_paper}. 
Additionally, we find the average orientation for Indian SNs to be $3.4218$ (Table~\ref{tab:network_data}), which aligns with the earlier reported value for the Asian SNs\cite{orientation_order_100_city}. 

\begin{table}[H]
    \centering
    
\resizebox{\textwidth}{!}{
\begin{tabular}{ |p{3.0cm}|p{3.0cm}|p{3.0cm}|p{3.0cm}|p{3.0cm}|p{3.0cm}| }

 \hline
 City &  \#Nodes & \#Edges & Orientation Entropy & $\alpha$ & $\rho$\\
 \hline
 Gurugram & 35580 & 47125 & 3.4152 & 0.1623 & 0.5519\\
 \hline
 Dehradun & 33117 & 38295 & 3.4544 & 0.0781 & 0.5663\\
 \hline
 Indore & 28998 & 40959 & 3.3302 & 0.2063 & 0.5241\\
 \hline
 Gandhinagar & 27672 & 35839 & 3.3812 & 0.1475 & 0.6466 \\
 \hline
 Gwalior & 26916 & 35612 & 3.4976 & 0.1616 & 0.5023\\
 \hline
 Srinagar & 26856 & 31687 & 3.5579 & 0.0899 & 0.5403\\
 \hline
 Jhansi & 25552 & 31875 & 3.5568 & 0.1238 & 0.6134\\
 \hline
 Jaipur & 25236 & 32276 & 3.3707 & 0.1395 & 0.7590\\
 \hline
 Madurai & 23670 & 32106 & 3.3330 & 0.1782 & 0.4847\\
 \hline
 Ambala & 23580 & 31359 & 3.3975 & 0.1649 & 0.6562\\
 \hline
 Thanjavur & 20986 & 27778 & 3.4003 & 0.1619 & 0.5892\\
 \hline
 Kollam & 20930 & 24474 & 3.4007 & 0.0847 & 0.4664\\
 \hline
 Patna & 19034 & 24436 & 3.2511 & 0.1419 & 0.5676\\
 \hline
 Kozhikode & 18204 & 22311 & 3.4844 & 0.1128 & 0.5060\\
 \hline
 New Delhi & 17016 & 23055 & 3.5109 & 0.1775 & 0.5772\\
 \hline
 Puducherry & 16683 & 22312 & 3.1617  & 0.1687 & 0.5038\\
 \hline
 Chandigarh & 15426 & 21663 & 2.9875 & 0.2020 & 0.5453\\
 \hline
 Vishakhapatnam & 14046 & 19662 & 3.3860 & 0.1999 & 0.5992\\
 \hline
 Ujjain & 9131 & 12663 & 3.3747 & 0.1935 & 0.5090\\
 \hline
 Imphal & 8691 & 10304 & 3.4648 & 0.0928 & 0.6181\\
 \hline
 Navi Mumbai & 8364 & 11654 & 3.4432 & 0.1967 & 0.5428\\
 \hline
 Ladakh & 8320 & 9962 & 3.5785 & 0.0987 & 0.5740\\
 \hline
 Tirupati & 7642 & 10562 & 3.1090 & 0.1911 & 0.5460\\
 \hline
 Aizwal & 6704 & 7749 & 3.5762 & 0.0780 & 0.5252\\
 \hline
 Shimla & 6560 & 7018 & 3.5800 & 0.0349 & 0.5896\\
 \hline
 Leh & 6246 & 7604 & 3.5759 & 0.1088 & 0.5824\\
 \hline
 Guwahati & 5338 & 6705 & 3.5055 & 0.1281 & 0.4752\\ 
 \hline
 SouthAndaman & 4144 & 4767 & 3.5768 & 0.0753 & 0.4332\\
 \hline
 Kochi & 4051 & 4852 & 3.1491 & 0.0990 & 0.4917\\
 \hline
 Kohima & 1782 & 2095 & 3.5737 & 0.0882 & 0.7114\\
 \hline
 Daman & 1203 & 1506 & 3.5023 & 0.1266 & 0.5222\\
 \hline
 Gangtok & 1130 & 1277 & 3.5562 &0.0656 & 0.5450\\
 \hline
 Dholera & 841 & 1183 & 3.4757 & 0.2045 & 0.6025\\
 \hline

\end{tabular} }
\caption{ {\bf Statistics of the different SNs studied:} This table summarizes the size (N), edge density ($E_d$), orientation Entropy, Meshedness coefficient ($\alpha$), and the weighted degree assortativity ($\rho$) of the studied SNs. All the studied SNs are sparse and coexist with the tree and lattice topology. The average Meshedness and orientation entropy are $0.1359$ and $3.4218$, respectively.}
    \label{tab:network_data}
\end{table}


\subsubsection{Path-Length Distribution of the streets}
\begin{figure}[htp!]
   \centering
   \includegraphics[width=0.76\textwidth]{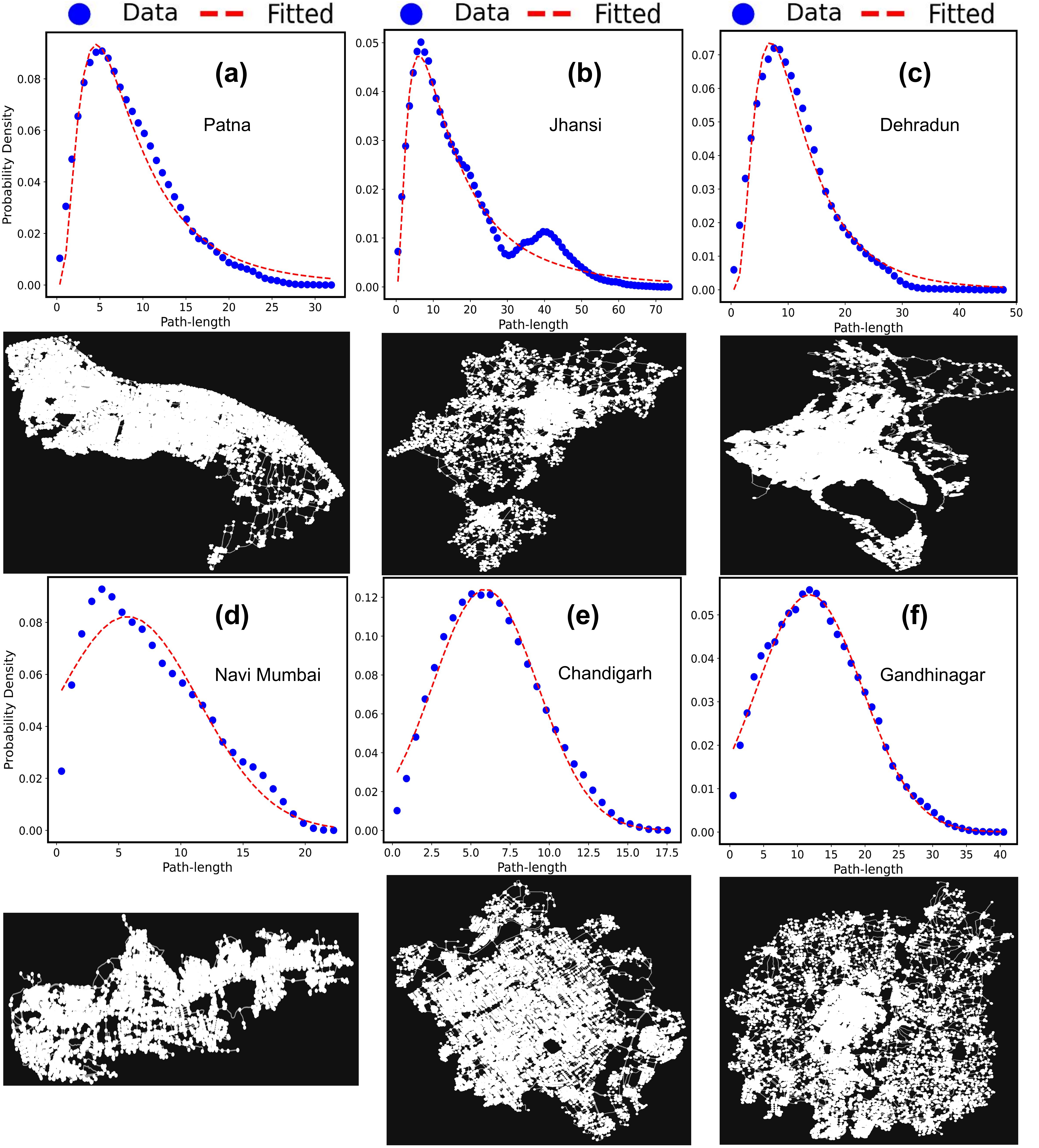}
    \caption{{\bf Path-Length Distribution of Organic and Planned SNs: }Sub-figures display the path-length distribution for various SNs, while the lower panel below them plots visualizations of the studied SNs. The path-length distribution for planned cities (d–f) closely follows a right-skewed normal distribution, whereas the distribution for organic and mixed cities aligns with a lognormal distribution.}
    \label{fig:Path_length_distribution}
\end{figure}

\noindent It is noted that the efficiency of the SNs discussed in the previous section is much higher than that of the corresponding degree-preserved random networks. We studied the path-length distribution (Methods) for all the SNs to elaborate on it. The SNs are embedded in space and, therefore, demand cost optimization along with higher clustering and efficiency. The cost associated with the SNs primarily comes from the length of the streets. Consequently, we calculate Dijkstra's path-length \cite{dijkstra_algorithm_original_paper} between all the pairs of nodes in an SN and, consequently, study their distribution. Dijkstra's algorithm provides the shortest paths between the nodes that minimize the associated cost, which in the SNs is the total distance traveled. 
Figs.\ref{fig:Path_length_distribution} (a-c, and d-f) illustrate the path-length distribution of various Indian cities. Similar plots for all the cities are available in the Supplementary material (Fig.~\ref{fig:supplementary_path_length_distribution}). We find that the path-length distribution of most of the planned cities resembles a right-skewed normal distribution and fits well with the function:
\begin{equation}
 F(x)=Ae^{-\frac{(x-\mu)^2}{2\sigma^2}} 
 \nonumber
 \end{equation}
where $\mu$ is the mean of the function and $\sigma$ is the standard deviation and $A$ is the normalizing constant. In contrast, the path-length distribution of most unplanned or mixed cities aligns more closely with a lognormal distribution and fits better with function: 
\begin{equation}
 G(x,s)=\frac{1}{sx\sqrt{2\pi}}e^{-\frac{\log^2{x}}{2s^2}} 
 \nonumber
 \end{equation}
 where 's' is the shape parameter. Such types of distributions have also been reported previously in the neuron length distribution \cite{barabasi_paper_physical_network_lognormal}. The right-skewed normal distribution of path lengths suggests that most path values cluster around a smaller value than the normal distributed network. This configuration indicates well-connected road networks, where most roads are easily accessible while only a few are difficult to reach.

The lognormal distribution of path lengths in most organic cities reveals a non-uniformity in the layout of streets. This distribution indicates that while the majority of pathways tend to be short, a significant tail of longer paths exists. Consequently, although most locations are well connected, the accessibility to certain areas is limited, making them more prominent compared to planned urban environments. Additionally, in these cities, roads are typically constructed in response to the evolving connectivity needs between various locations over time, contributing to their organic development.

Although we can distinguish certain cities based on their path-length distribution, generalizing this concept to differentiate organic cities from planned ones proves challenging. It is possible that a city was initially developed in a planned manner; however, over the years, the densification of the city may have made the original planning almost irrelevant \cite{entropy_40_cities}. Consequently, other factors that address the local behavior of streets will play a crucial role in understanding the planned versus unplanned nature of the SNs in these cities.

Additionally, we observed that some cities exhibit bimodal behavior in their path-length distribution, indicated by two distinct peaks. This phenomenon could be attributed to the presence of clusters within the cities, where the clusters are well-connected internally, but there are few connections between them, leading to the observed bimodal distribution.

\begin{figure}[htp!]
    \centering
    \includegraphics[width=0.7\textwidth]{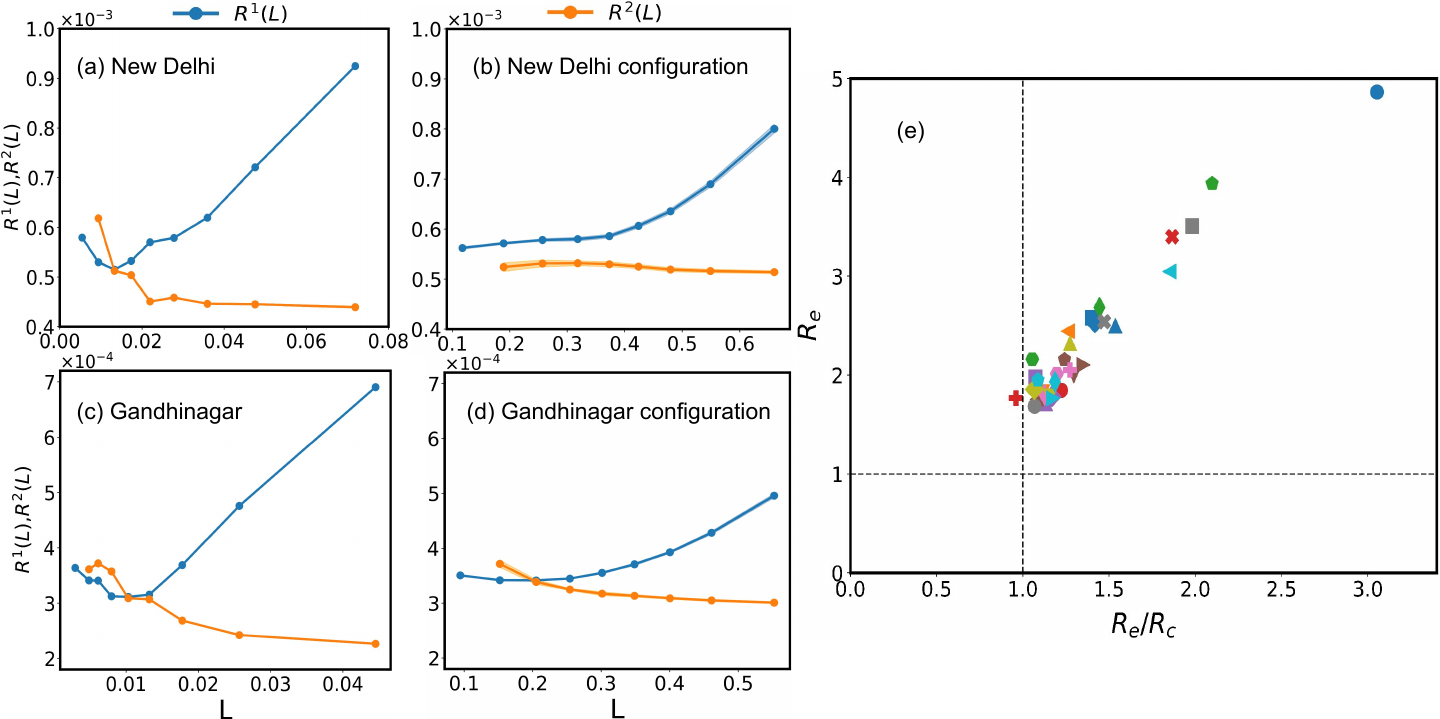}
    \caption{{\bf Rich Club of the long streets in SNs:} In this plot, we divide the entire SN into different bins, each with the same number of streets in ascending order of their lengths. To calculate $R^1(L)$, we count the number of streets that are connected in this bin and the number of connections the streets of this bin have with the streets of higher bins and normalize this count by the total number of streets in both this bin and the higher bins. 
For the calculation of  $R^2(L)$, we count the number of streets in the bins lower than this bin that is connected to the street of this bin and normalize it with the total number of streets in the bins lower than this bin. (Normalized length in a,b,c,d). The subplot (e) displays $R_e = \frac{R^1(L)}{ R^2(L)}$ for the empirical SN and its ratio with the corresponding degree preserved random network ($R_c$) for the last bin of all $33$ cities.}
    \label{fig:rcc_empirical}
\end{figure}

\subsection{Rich-club of long streets in SNs:}

We find moderate positive correlations in the weighted degree (see Table~\ref{tab:network_data}), indicating a preferential connection between the junctions per the number/length of the streets passing through them.

{\noindent} To explore this further, we introduce two metrics parameterized by the lengths of street segments, offering a closer look at the interplay between topology and spatial constraints.
We divided the total number of streets into different bins based on the increasing order of their length, ensuring that each bin contains an equal number of streets. The length of the bin, denoted as $\Delta L$, is defined by $L_2 - L_1$, where $L_1$ is the minimum length, and $L_2$ is the maximum street length in that bin. The first metric in each bin is defined as follows: 
 \begin{equation}
 R^1(L) = \frac{N_{S> L_1}}{TN_{S> L_1}}, 
 \nonumber
 \end{equation}
In the above equation, $N_{S> L_1}$ is the number of the neighboring streets of the streets with the length lying in the bin $\Delta L$ with the length greater than or equal to $L_1$, and $TN_{S> L_1}$ is the total number of streets of a length greater than or equal to $L_1$. 
This metric assesses the connectivity between streets of similar and those of longer lengths.     
The other metric we define to measure the connectivity of a street with the shorter streets is given by: 
\begin{equation}
R^2(L) = \frac{N_{S<L_1}}{TN_{S< L_1}}, 
\nonumber 
 \end{equation}
Here, $N_{S< L_1}$ is the number of neighboring streets of the streets in $\Delta L$ that are shorter than $L_1$, while  $TN_{S<L_1}$ is the total number of the streets that are shorter than $L_1$.
 Figs.~\ref{fig:rcc_empirical}(a) and (c) plots $R^1(L)$ (blue circles)  and $R^2(L)$ (orange circles) values for the SNs of New Delhi and Gandhinagar. For New Delhi, the value of $R^2(L)$ is higher than the $R^1(L)$ for the lowest bin and decreases for the higher lengths. Meanwhile,  the value of $R^1(L)$ first decreases and then increases rapidly (Fig.~\ref{fig:rcc_empirical}(a)),  indicating shorter streets better connect with the shorter streets, whereas the longer streets better connect with the streets of similar or higher lengths. Similarly, Gandhinagar's shorter streets connect better with the shorter streets, and the longer streets connect more closely with similar or longer streets. 
 Furthermore, to avoid the cases of random occurrence of better connectivity among the long streets, we do the same study for the corresponding degree preserved random networks, as plotted in Figs.~\ref{fig:rcc_empirical}(b) and (d). We find that in the random networks, too, for the prolonged streets, show $R^1(L)>R^2(L)$, however not as much in strength as the empirical networks. To elaborate this we calculate the ratio  $R_e = \frac{R^1(L)}{ R^2(L)}$ for the last bin of the empirical networks and similarly $R_c =  \frac{R^1_{c}(L)}{ R^2_{c}(L)}$ for the corresponding degree preserved random networks. 
 
 $R_e >1$ shows that the streets in the last bin connect more to the streets of similar lengths, whereas the $R_e <1$ implies that streets in the last bin connect more to the streets smaller than them. Also, the co-existence of $R_e>1$ and $\ \frac {R_e}{R_c}>1$ ensures the non-random occurrence of better connectivity of longer streets with other streets with similar lengths. Figs.~\ref{fig:rcc_empirical}(a-d) depict that this property is valid for the SNs of New Delhi and Gandhinagar.  
 
Further, to demonstrate the consistency of the higher connectivity among the longer streets across all studied cities, we calculated $R_e$ and $R_e/R_c$ for the last bin in each studied city. The results are plotted in Fig. ~\ref{fig:rcc_empirical} (e), indicating that, except for Tirupati, both values are more significant than $1$ for all studied cities. 
A similar trend of a higher connection among the high-degree nodes than a corresponding random network has previously been observed in the brain networks \cite{rich_club_paper_brain_network, prateek_paper_brain_network}, known as the Rich-club phenomenon. Therefore, we state that this phenomenon of longer streets connecting more with similar or longer streets is the rich-club phenomenon in street networks.

\begin{figure}[htp!]
    \centering
    \includegraphics[width=0.46\linewidth]{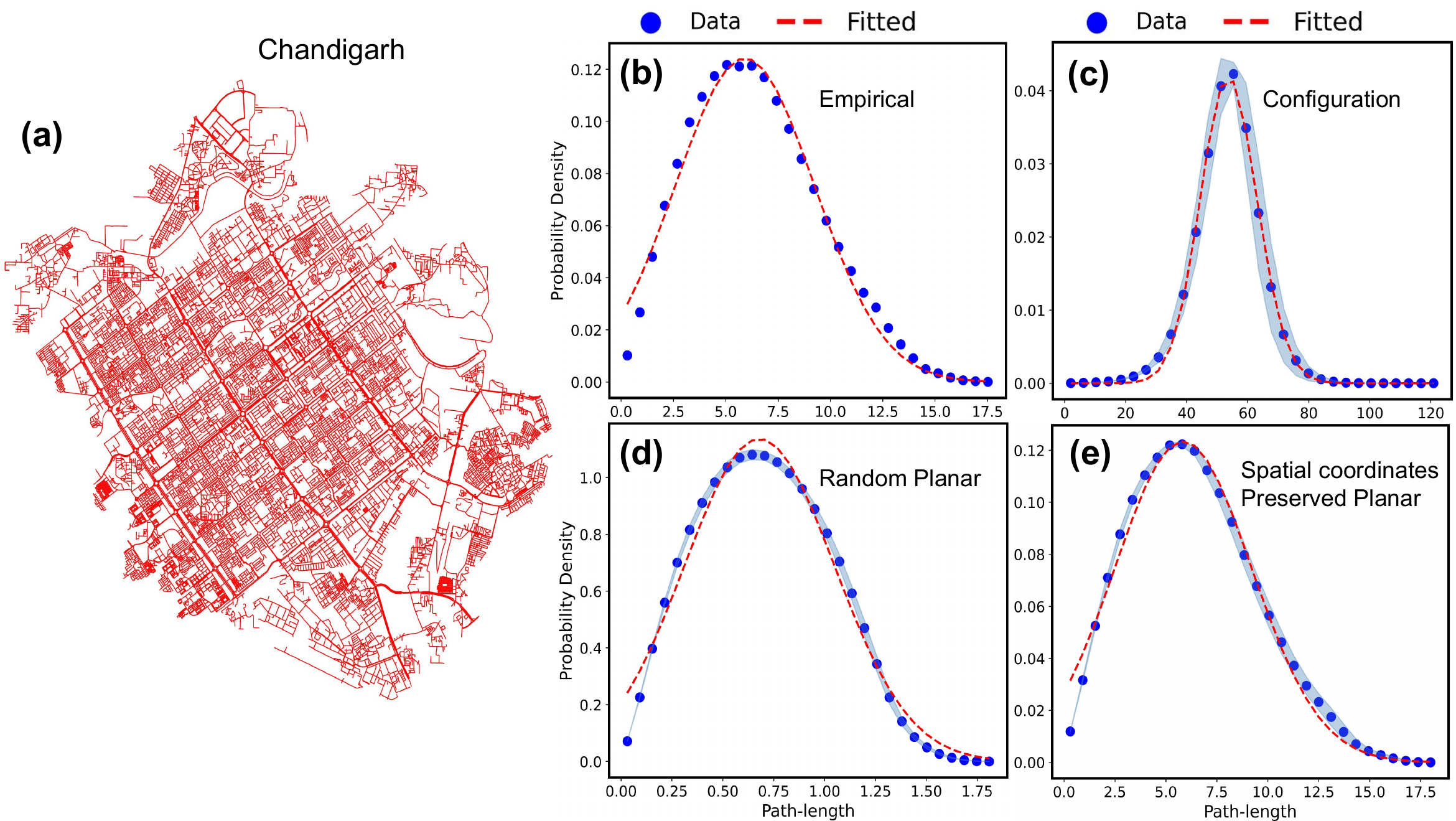}
    \includegraphics[width=0.46\linewidth]{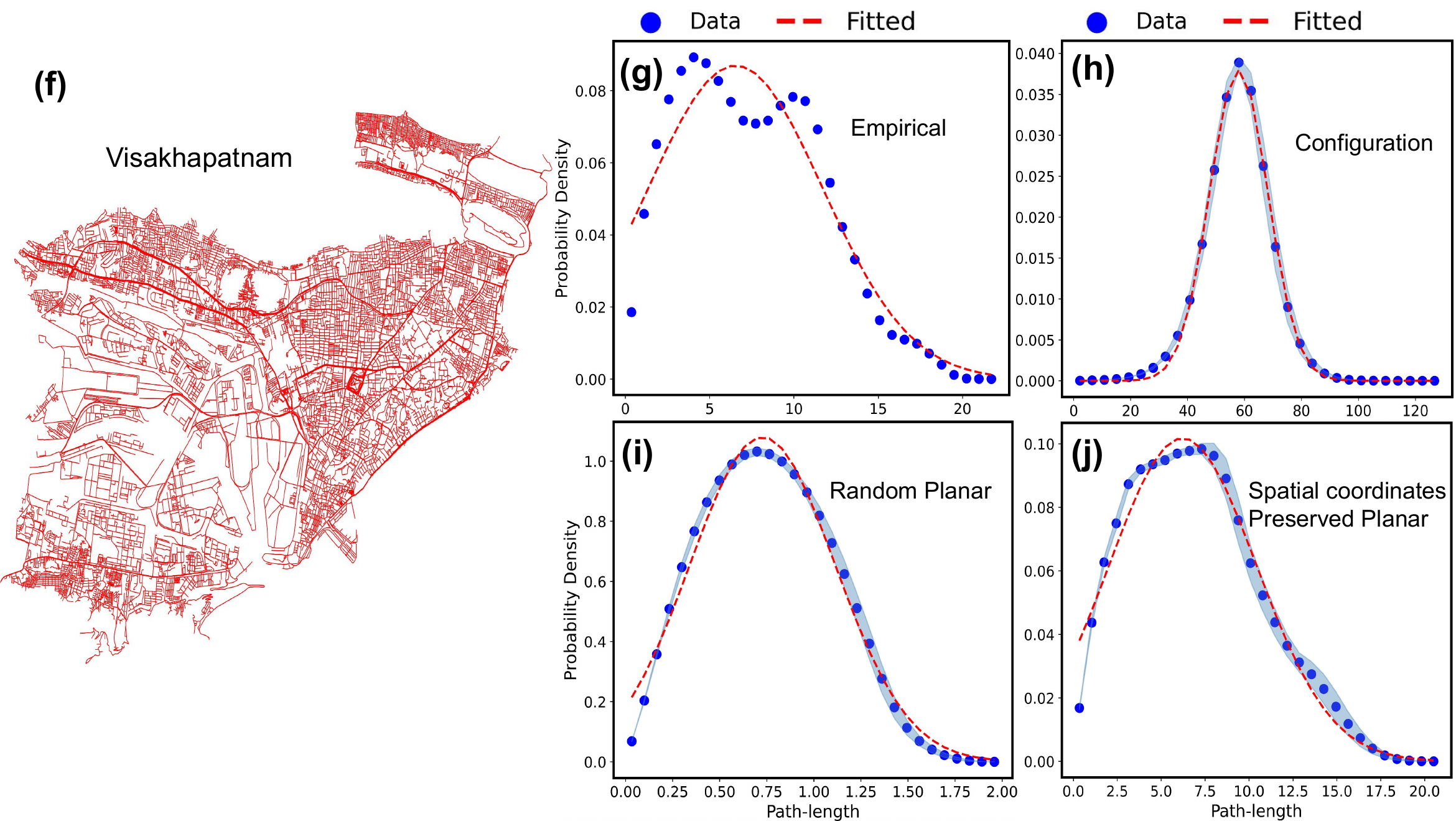}
\caption{{\bf Comparison of the path-length Distribution of the empirical SNs with the model networks:} Subfigs. (a) and (f) illustrate the street layouts of the cities of Chandigarh and Vishakhapatnam. Subfigs. (b-e) present the path-length distributions for the empirical network of Chandigarh, a corresponding degree-preserved random network, a random planar network, and a coordinates-preserved planar network, respectively. Similarly, subfigs. (g-j) depict the same for Vishakhapatnam. This analysis includes the results for $10$ random planar graphs.}
    \label{fig:Path_length_distribution_randomplanar}
\end{figure}
\begin{figure}[htp!]
    \centering
    \includegraphics[width=0.65\linewidth]{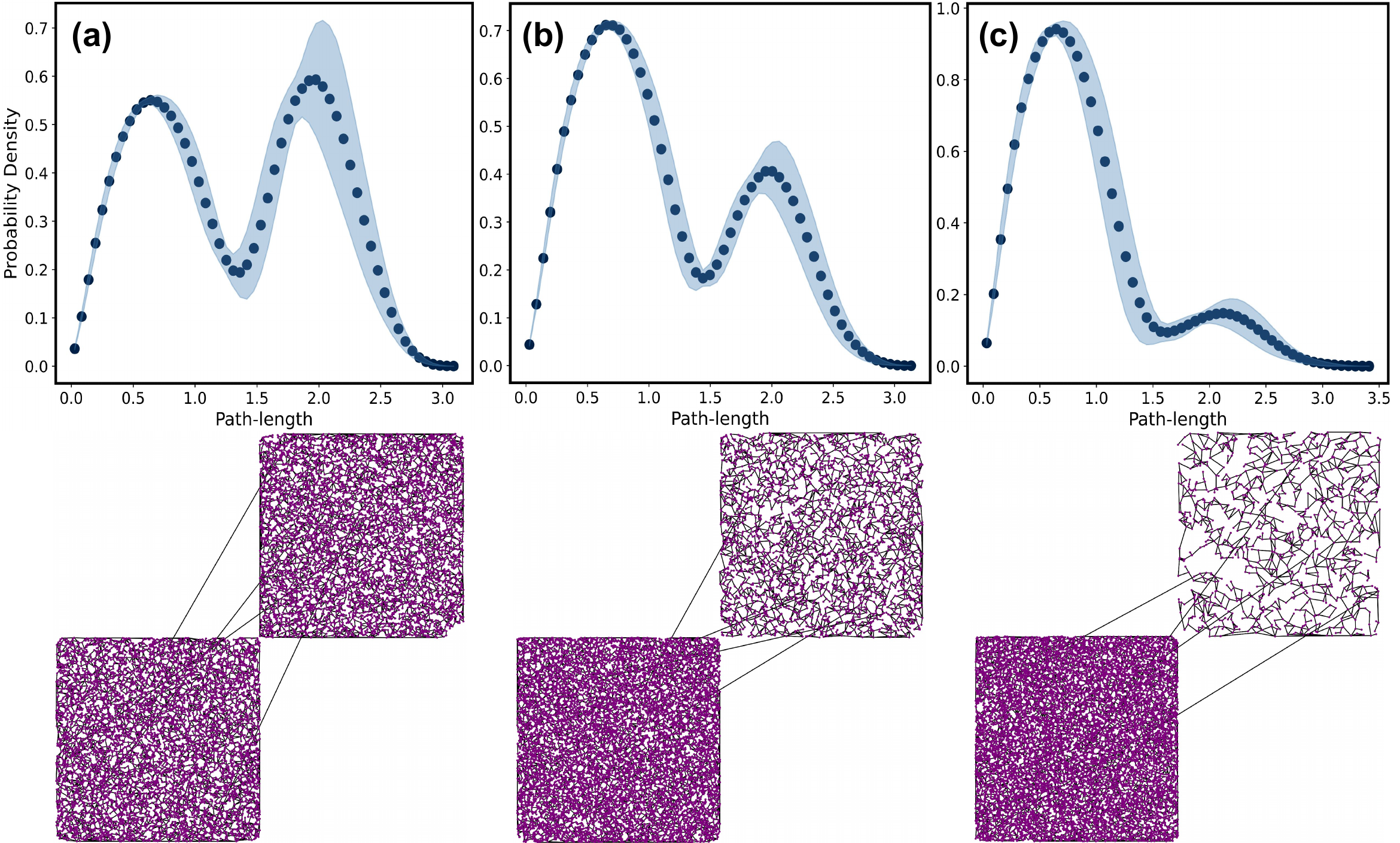}
    \caption{{\bf Origin of the bimodal path-length distribution in a clustered random planar network}: The clustered planar networks, where the two random planar networks of various sizes are connected with four edges. Figs (a-c) plot the path-length distribution for the clustered random planar network of sizes $N_1, \; N_2 = 7000, 7000 $, $N_1, \; N_2 = 11000, 3000 $, and $N_1, \; N_2 = 13000, 1000 $ respectively. The plots represent averages from 10 random networks, with the blue paths surrounding the scattered points indicating the standard deviation.
    Plots below sub-figs (a-c) present one of the visualizations of the corresponding network.}
    \label{fig:clustered_random_planarnetwors}
\end{figure}
\begin{figure}[htp!]
    \centering
    \includegraphics[width=0.76\linewidth]{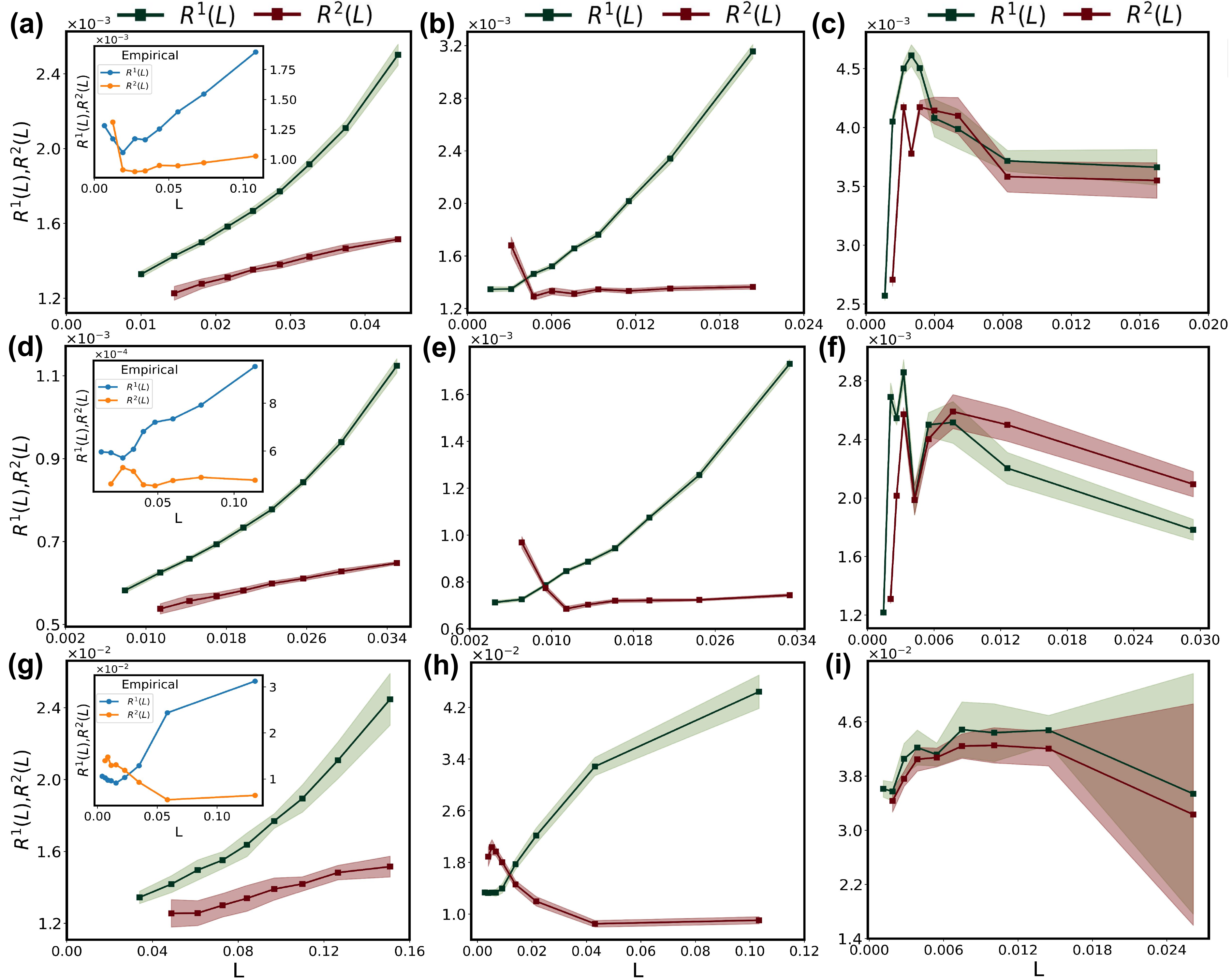}
    \caption{{\bf Rich Club for planar network }: Subplots (a-c) display the variation of parameters $R^1(L)$ and $R^2(L)$ with the length $L$ of the streets for the Navi Mumbai (top: a-c), Chandigarh (middle: d-f), and Dholera (bottom: g-i). 
The subplots (a, d, g) correspond to the random planar networks, with the inset displaying empirical SNs. Subplots (b, e, h) are for the position coordinate preserved planar networks, and subplots (c, f, i) correspond to the model spatial small-world network, as discussed in the results section. 
The ratio of $R_1$ and $R_2$  for the empirical networks, random planar networks, coordinate preserved planar networks, and model spatial small-world networks are denoted as $R_e$, $R_p$, $R_{sp}$, and $R_m$, respectively. We find that for the Navi Mumbai, Chandigarh, and Dholera, these values are as follows: $R_e = 1.85, \; 1.997, \; 4.89  $, $R_p =1.65, \;1.73, \; 1.61 $, $R_{sp} = 2.31, \; 2.32, \; 4.92$, and $R_m = 1.03, 0.85, 1.09 $,  respectively.}
    \label{fig:RCC-random-planar-Spatial-SW-model}
\end{figure}
\subsection{A comparison with the model networks:}
\noindent We find that all the studied SNs, except Dholera, Daman, Kohima, South Andaman, and Imphal, violated the condition of being strictly planar(Table: \ref{tab:planarity_data}). This non-planarity allows the edges to cross each other,  introducing a level of randomness that deviates from the strict requirements of planarity. The non-planarity observed in the current SNs could be attributed to the presence of the flyovers and tunnels\cite{planar_and_non_planar_models_street_network}. Nonetheless,  it is essential to note that the 
 number of edges that induce the non-planar behavior in the network remains at or below $1\%$ of the total edges of the network, and the majority of the network retains planarity. 

\noindent We compare the empirical SNs with the random configuration and the random planar networks of the exact sizes. We use the Delaunay triangulation(DT) algorithm\cite{deluany_triangulation_algorithm_scipy_paper} to generate the random planar graphs. We generate the random planar graphs of the size of the empirical networks by considering the random distribution of the points on a square of unit dimensions using the DT algorithm. Then, we remove the edges randomly until their number is similar to those in the empirical networks. Additionally, we generate the planar graphs by preserving the position coordinates of the empirical networks using the DT algorithm. 

\noindent Moreover, we also compare the results of the empirical networks with the spatial small-world network using the model as discussed follows: 
The higher clustering and efficiency of the empirical network is modeled by the is characterized by Watts-Strogatz networks \cite{strogats_small_world}. The unique combination of high local clustering and short average path lengths makes them practical for modeling various real-world systems. The distance distribution is assumed to be uniform in the original Watts-Strogatz model\cite{strogats_small_world}. However, in most real-world systems, the distance between nodes is not uniform, and in street networks, there are way more short connections than long connections\cite{3_cities_parameteric_topological_study}, so we generate the spatial small-world networks with the following model:
\begin{equation}
P_{ij} \propto d_{ij}^{-\beta}, 
\nonumber
\end{equation}

where $P_{ij}$ is the probability of the connection between junctions $i$ and $j$, and $d_{ij}$ is the distance between these junctions. The exponent $\beta$ controls the proportion of the long connections. For $\beta=0$, the length distribution is uniform, and the network is an Erd\H os–R\'enyi random network. As $\beta$ increases, the network transitions into a spatial small-world structure\cite{phase_transition_spatial_small_world, spatial_small_world}. The length distribution enters the scale-free regime for the values of the $\beta$ in $2$ and $3$. At very high values of the $\beta$, the fraction of the long connections decreases, resulting in a network that resembles a lattice. The deviation from uniformity provides valuable insights into changes in the topology. To show that the connectivity pattern of the empirical SNs is not associated with the small-world property of the SNs, it is crucial to explore the connectivity patterns of these networks, too. In generating the spatial small-world model networks as defined above, we keep the position coordinates of the junction the same as for the empirical networks.

\subsubsection{Comparison of the path-length distribution}
\noindent Fig.~(\ref{fig:Path_length_distribution_randomplanar}) plots the path-length distribution of the empirical SNs of Chandigarh and Vishakhapatnam and the corresponding model networks. 
The path lengths in the empirical SN of Chandigarh exhibit a right-skewed normal distribution ( Fig.~(\ref{fig:Path_length_distribution_randomplanar}(b)). The path-length distribution for the corresponding random configuration network is normally distributed but has a higher average than that of the empirical networks ( Fig.~(\ref{fig:Path_length_distribution_randomplanar}(c)). 
The path-length distribution of the random planar network closely resembles that of the empirical network, whereas the path-length distribution of the spatial cor-ordinate preserved planar network completely aligns with the empirical network (Fig.~\ref{fig:Path_length_distribution_randomplanar}(d, e)). 
Furthermore, the path-length distribution for the Vishakhapatnam shows a bimodal distribution, and both the corresponding configuration and the random planar network do not show this behavior (Figs.~\ref{fig:Path_length_distribution_randomplanar}(g-i)). However,  a planar network generated preserving the spatial coordinates of the empirical network further replicates the bimodal behaviour of the path-length distribution (Figs.~\ref{fig:Path_length_distribution_randomplanar}(j)). 

A bimodal path-length distribution for the city of Vishakhapatnam suggests the presence of two distinct clusters within the city. Therefore, a planar network generated without accounting for these two clusters results in an unimodal path-length distribution (Fig. \ref{fig:Path_length_distribution_randomplanar}(i)). To further demonstrate that two clusters within a city create a bimodal distribution of path lengths, we generate a random planar network consisting of land areas of varying sizes connected by four links. We then analyze the resulting path length distribution. Our findings indicate that these networks exhibit bimodal path length distributions, with the heights of the peaks varying based on the size difference between the two clusters (see Fig~\ref{fig:clustered_random_planarnetwors}). The left peak, associated with lower path length values, represents communication between junctions within the same cluster. In contrast, the right peak, corresponding to higher path length values, reflects communication between the two clusters. Many peaks can be similarly seen in a network with multiple clusters. 

\subsubsection{Comparison of the length-based connectivity}
This section compares the length-based connectivity patterns observed in the empirical networks with those from the corresponding model networks. Figures \ref{fig:RCC-random-planar-Spatial-SW-model} (a-f) illustrate the \( R^1(L) \) and \( R^2(L) \) values concerning the length \( L \) for random-planar networks, spatial embedding preserving planar networks, and model-generated spatial small-world networks associated with the street networks (SNs) of Navi Mumbai and Chandigarh. The insets in Figures \ref{fig:RCC-random-planar-Spatial-SW-model} (a) and (d) display the empirical SNs. As discussed in the previous section, the empirical SNs demonstrate improved connectivity among shorter streets of similar lengths, as well as enhanced connectivity among longer streets that share similar characteristics; for the final bin, the ratio \( R_e = R^1(L)/R^2(L) \) is 1.8469 for Navi Mumbai and 1.9975 for Chandigarh. In contrast, the random planar networks do not exhibit superior connectivity among smaller streets but show increased connectivity among longer streets. In this case, we find that for the last bin, the ratio \( R_p = R^1(L)/R^2(L) \) equals 1.65 and 1.73 for Navi Mumbai and Chandigarh, respectively, indicating that although this phenomenon is present, it occurs to a lesser degree than in the empirical networks.

\noindent The planar networks generated while maintaining the spatial embedding replicate a greater connectivity of smaller and longer streets. The ratio of $R^1(L)$ to $R^2(L)$ for the final bin, $R_{sp} = 2.31$ for Navi Mumbai and $R_{sp} = 2.32$ for Chandigarh, indicates that this phenomenon is even more pronounced in these instances.

Additionally, we construct the corresponding spatial small-world networks for Dholera, Navi Mumbai, and Chandigarh with parameters set at $\beta = 1.4, \; 2.6,$ and $2.8$, respectively, while maintaining the spatial coordinates. However, these spatial small-world networks do not replicate the connectivity phenomena exhibited by the empirical street networks, and the connectivity of the streets remains unaffected by their sizes (see sub-figs.\ref{fig:RCC-random-planar-Spatial-SW-model} (c, f, i)).

It is important to note that empirical SNs for Navi Mumbai and Chandigarh are not strictly planar, whereas the planar networks generated preserving spatial embedding are strictly planar; a slightly enhanced connectivity of the longer streets than that displayed by the empirical networks could be attributed to it. The observation that the SNs of Dholera city, which is strictly planar, have similar connectivity to the longer streets as corresponding spatial embedding preserved planar networks confirms this finding (Fig.\ref{fig:RCC-random-planar-Spatial-SW-model}). Therefore, planarity and spatial embedding give rise to the length-based connectivity between the streets. 

\subsection{Robustness analysis of street networks under random errors and targeted attacks:}
\noindent
We further investigate the robustness of the SNs to check their ability to remain functional despite various disruptions. Accidentally/targeted damage to a junction or a street may completely disturb/delay the traffic dynamics. Also, it may cause a cascading effect on the city's overall functionality. Errors may arise from random damage to any street/junction. At the same time, attacks can be strategically aimed at the most critical streets/junctions identified by their betweenness centrality value. 
 
\begin{figure}[htp!]
    \centering
    \includegraphics[width=0.650\textwidth]{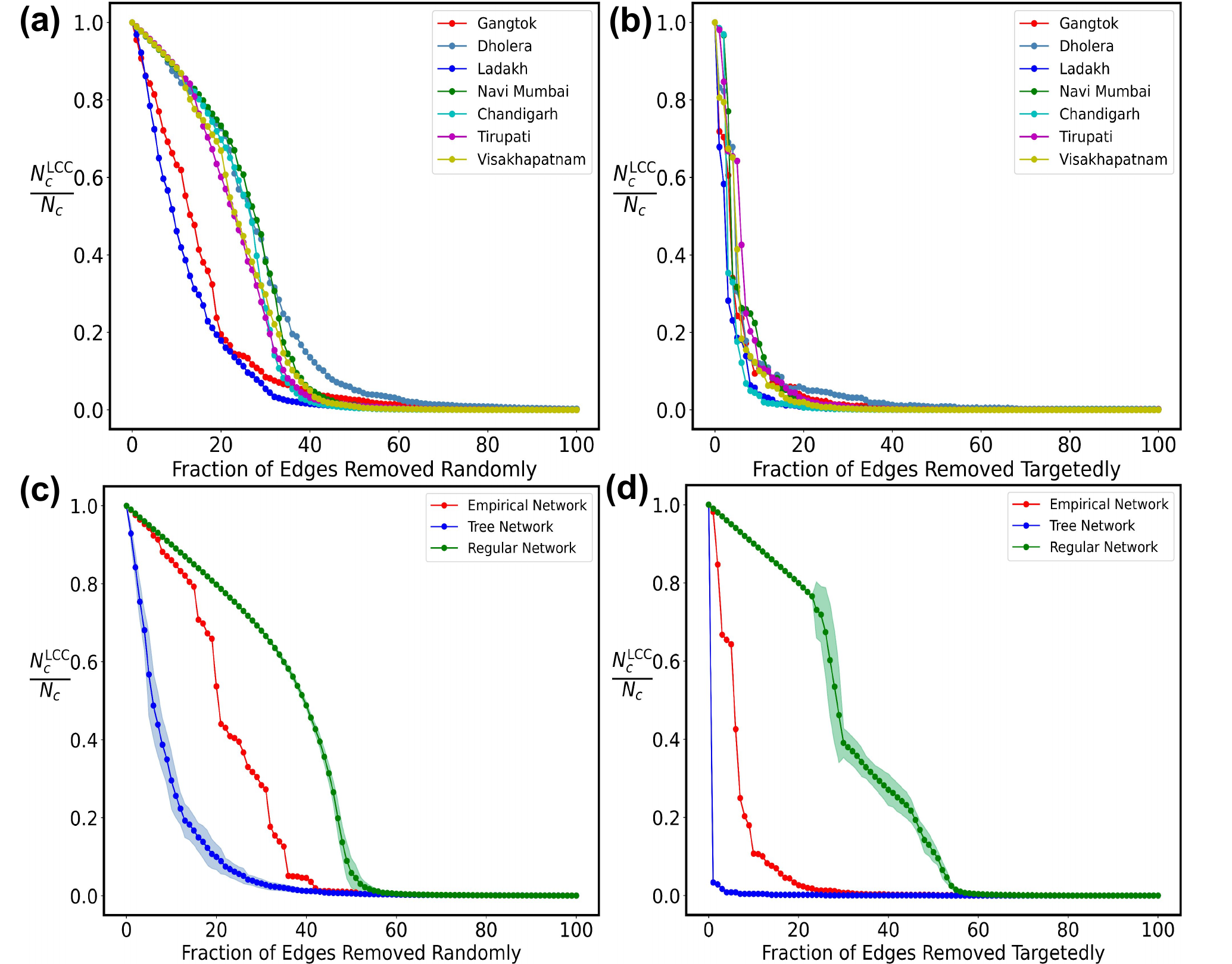}
    \caption{{\bf Robustness of the SNs:}  The subplots (a-d) display the fraction of the nodes in the largest connected component (LCC) concerning random removals (a, c) and the edge betweenness centrality based removals (b, d) of the streets. Subplots (a) and (b) present the robustness of the cities Gangtok, Dholera, Ladakh, Navi Mumbai, Chandigarh, Tirupati, and Vishakhapatnam. Meanwhile, the subplots (c) and (d) compare the robustness of Tirupati city with the corresponding Cayley tree networks and the regular random networks. The observation shows that when faced with the attacks, the empirical networks show behavior close to that of the tree networks, indicating their vulnerability. }
    \label{fig:robustness-empirical}
\end{figure}

Our study examined the SNs' robustness in responding to random errors and targeted attacks on the streets.
Fig.\ref{fig:robustness-empirical}(a) plots the fraction of nodes in the most significant connected component (LCC) concerning the fraction of the randomly removed streets for seven SNs. We see an impact of the latitude on the robustness; the hilly cities Gangtok and Laddakh are less robust than the other cities. However, this phenomenon is not present for the betweenness centrality-based attacks plotted in Fig.\ref{fig:robustness-empirical}(b), and all the studied cities are fragile and collapse for the removal of only $10\%$ of the streets. 

Furthermore, we also compared the robustness of the empirical street networks with tree and 2-D lattice networks, as shown in  Fig.\ref{fig:robustness-empirical}(c-d). We find that for both the errors and the attacks, the empirical street network lies somewhere between the robustness of the tree and the lattice network. The lattice networks are more resilient to errors and attacks than the empirical and the tree networks. The empirical networks are as fragile as the tree networks against the betweenness centrality-based attacks. This also indicates that an enhancement in the latticization may further improve the robustness of the Indian SNs.

\section{Conclusion and Discussion:}
\noindent  
 The topology of SNs matters as they influence the socioeconomic growth of a city by regulating the flow of people/goods. Our study examines the topology of the Indian SNs and the role of physical characteristics in shaping them. We find that all the studied networks have the spatial small-world property, with the efficiency of the empirical SNs being much higher than the corresponding degree preserved random networks. It suggests that most roads
are easily accessible by traversing only a few connections, highlighting the network's effectiveness. The spatial-small world organization promises a network to provide maximum efficiency at a
minimum cost. For a developing country like India, cost and efficiency should be the cause of concern, and the spatial small-world feature ensures that. Moreover, the coexistence of the SN's lattice and tree features ensures both pedestrian safety and better connectivity.
To explore more about the cause of the higher efficiency of the SNs, we study the path-length distribution and find that all studied SNs' Dijkstra's path-length distributions fit well with either the right skewed normal distribution or a lognormal distribution. The right skewness of the path-length distribution indicates that most of the junctions in the SNs can be covered in smaller steps than the corresponding random graphs. Also, a longer tail in the empirical SNs indicates that reaching some remote locations may require navigating multiple paths, diminishing their accessibility and putting the network at risk of attacks. Moreover, the observance of the bimodal distribution for the path length indicates the existence of the clustered city. 
No correlation between the network size and the average path length of the cities indicates that the path length primarily depends on the topology of the SNs rather than their respective sizes. We also compare the path-length distribution of the empirical SNs with the random planar networks and the spatial coordinates preserving planar networks. The path-length distribution of random planar networks and planar networks generated while preserving spatial coordinates aligns with the empirical skewed path-length distribution. This indicates that planarity is critical in shaping the topology of SNs and, hence, traffic dynamics.

\noindent  Furthermore, we find that while the planarity explains the skewed path-length distributions, the topology of the empirical SNs exhibits additional structural nuances. The streets tend to connect more with the streets of similar lengths; the smaller streets connect more to the smaller
streets, and longer streets predominantly link with the other similar or longer streets. The connectivity of the longer streets with those of similar or longer streets reveals the existence of
the Rich-club phenomenon in the street networks. A comparison of the empirical SNs with the corresponding degree
preserved Random networks, the model network generated considering the length distribution being power-law, and
the random planar networks further confirm the non-random existence of the rich-club phenomenon. Additionally, spatial coordinates preserving planar networks mimic both: the better connection between smaller and longer streets, as shown by the empirical networks, highlighting how junction positioning and planarity shape the topology of SNs. 
Moreover, Indian SNs are more fragile regarding the attacks than the previously studied Zurich street network \cite{robustness_zurich_road_network}. This suggests that SNs of  Indian cities lack resilience against attacks,  hinting at the need for specific structural changes to make them more robust. Also, the robustness of all the cities studied depends on geography, such as latitude, and the network size has no significant role.

\noindent  These results indicate that though spatial small-worldness is the property of street networks, its topology differs due to the restriction from its physicality. Also, it is insufficient to state that the SNs are merely planar; instead, they exhibit a more complex topology where the geography and spatial arrangement of the junctions significantly shape street connectivity and stability.

\section{Methods} 
{\bf Clustering and Efficiency:} We calculated the weighted clustering coefficient as follows \cite{weighted_clustering_nx_paper}: \begin{equation} \overset{\sim}{C_{i}} = \frac{2}{k_i (k_i - 1)} \sum_{j,k}(\overset{\sim}{w_{ij}}\overset{\sim}{w_{jk}}\overset{\sim}{w_{ki}})^{1/3}, \nonumber \end{equation} where $\overset{\sim}{w_{ij}} = \frac{w_{ij}}{max(w_{ij})}$ and $w_{ij}$ represents the length of the street connecting junctions (nodes) $i$ and $j$. In this equation, $k_i$ denotes the degree of junction (node) $i$.  In this case, the weights are the lengths of the streets, and this metric also offers insights into spatial clustering. Furthermore, the efficiency of a network indicates how well the nodes can communicate with one another \cite{dual_graph_main_paper,efficiency_first_paper}. It is calculated as follows: \begin{equation} E(G) = \frac{1}{N(N-1)}\sum \frac{1}{d_{ij}}, \nonumber \end{equation} where $N$ is the number of nodes and $d_{ij}$ is the shortest path between nodes $i$ and $j$. The shortest distance is determined using Dijkstra's algorithm \cite{dijkstra_algorithm_original_paper}, taking the spatial aspect of street networks (SNs) into account.

{\bf Weighted degree assortativity:} It gives insight into the propensity of nodes to connect with similar nodes of similar weighted degrees. It can be calculated using the Pearson correlation coefficient \cite{assortativity_mixing}. 
\newline 
{\bf Meshedness Coefficient:} The Meshedness Coefficient quantifies the ratio of the number of loops in a network to the maximum number of possible loops. It is defined as follows: 
\begin{equation} \alpha = \frac{u}{v} = \frac{k - n + 1}{2n - 5}, \nonumber \end{equation} 
where $\alpha$ represents the Meshedness coefficient, $k$ is the total number of edges, $n$ is the number of nodes and $u$ and $v$ are the number of existing loops and the maximum possible loops, respectively, in the network \cite{meshedness_paper, book_transport_systems}. 
\newline 
{\bf Orientation Entropy and Orientation Order}: We analyze the Shannon entropy \cite{shannon_entropy} of the distribution of street orientations. The Shannon Entropy for a probability distribution is expressed as: \begin{equation} H_O = -k \sum_{i=1}^{n} P(O_i) \log(P(O_i)) \nonumber \end{equation} 
where $n$ is the number of bins, and $P(O_i)$ is the probability of a street's orientation falling within the $i^{th}$ bin. In the continuation of the same approach, a related parameter known as orientation order has been defined to provide insight into the orderliness of the street network \cite{orientation_order_100_city}. This parameter helps to categorize whether a street network (SN) is ordered. The orientation order is calculated as:
 \begin{equation} \phi = 1 - \left( \frac{H_O - H_g}{H_{\text{max}} - H_g} \right)^2 \nonumber \end{equation} 
where $H_O$ is the orientation entropy, $H_g$ is the orientation of a perfect grid network. In our case, $H_g = 1.386$ nats and $H_{\text{max}} = 3.584$ nats, with $H_{\text{max}}$ representing the maximum possible entropy for the most disordered network. This maximum entropy corresponds to a network with an equal distribution of streets in all directions \cite{orientation_order_100_city}.

\section*{Acknowledgments}
We thank the Department of Science and Technology (DST), Government of India, for their financial support through the DST INSPIRE Faculty grant. We also thank IISER Tirupati for providing the necessary infrastructure and the High-Performance Computing (HPC) facility. Additionally, RK appreciates the input and assistance from lab members at IISER Tirupati and friends, especially  Dhruvi Panchal, for their valuable contributions.

\bibliographystyle{unsrt}  
\bibliography{references}  

\newpage

\section*{Supplementary Information}

\setcounter{figure}{0}
\renewcommand{\thefigure}{S\arabic{figure}}
\setcounter{table}{0}
\renewcommand{\thetable}{S\arabic{table}}

{\bf Details of the Data Extraction and the path-length distribution study:}

In extracting the data from OpenStreetMap (OSM), we used the function $graph\_from\_place$ in OSMnx, where the place is the name of the cities we wanted to extract the data. As we aim to study all the streets, we opted for $network\_type = all$. 
Next, we relabel the nodes, start the node label from 0, and extract area details using the $geocode\_to\_gdf$ function in OSMnx. The output of this function is an unweighted MultiDiGraph, which we converted into an undirected network using the NetworkX Graph function.
We subsequently assigned weights to the edges based on the geodesic distance between the junctions. We used the `geodesic' function to calculate these distances from the  `$geopy.distance$' library, measuring the Geodesic distance in kilometers using the nodes' coordinates(latitude and longitude). We further assigned the weights as the edge attributes of the network using Networkx, removed the self-loops if any were present, and saved the network as the Graphml file using NetworkX. 

Additionally, the path lengths of the city's street network, as illustrated in Fig.\ref{fig:Path_length_distribution}, were calculated using the well-established  Dijkstra's algorithm. Dijkstra's algorithm helps find the shortest and cheapest route between various points in a network. 

\begin{figure}[htp!]
    \centering
    \includegraphics[width=1\linewidth]{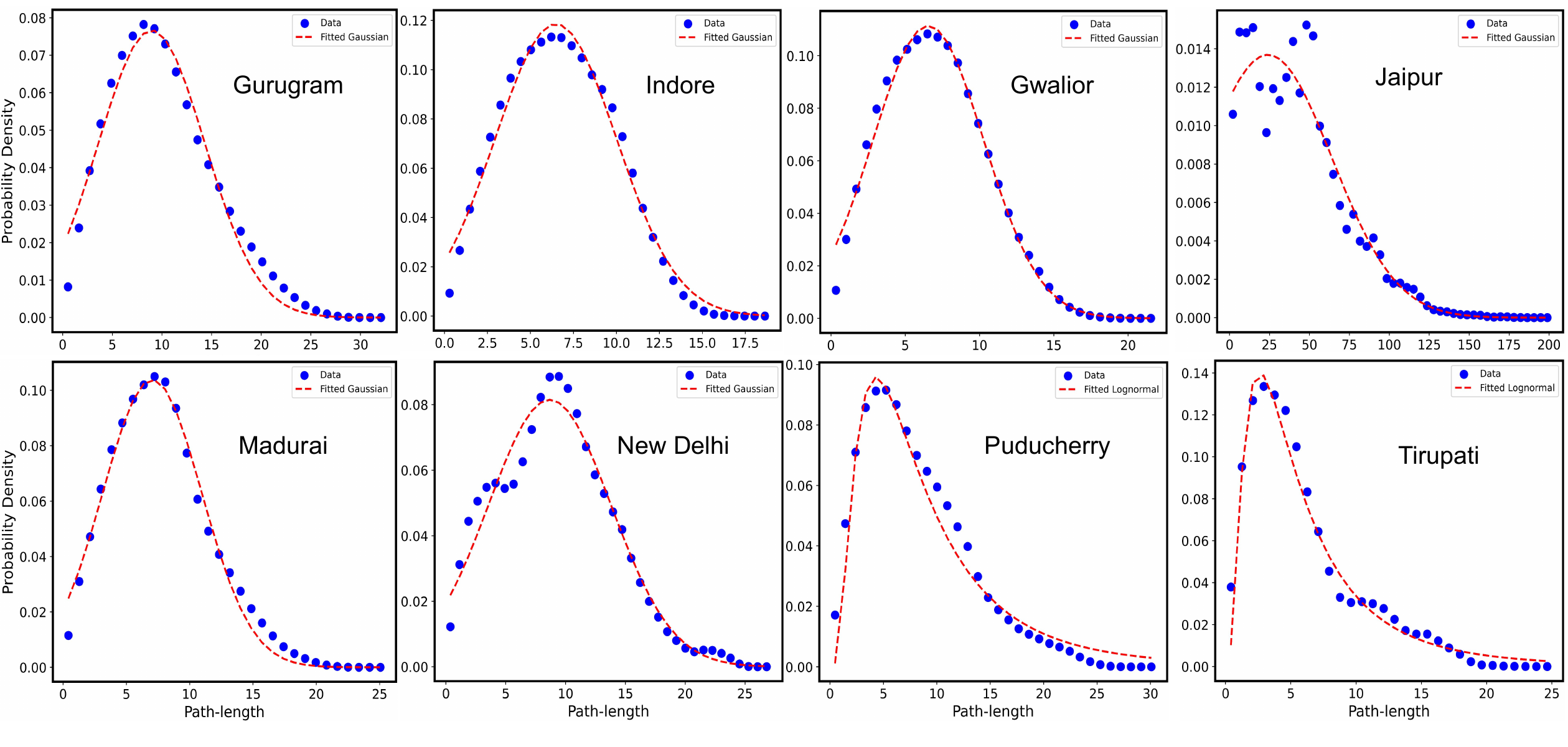}
    \caption{{\bf Path-Length Distribution: } This figure illustrates the Dijkstra's path length distribution for several other cities. It highlights the consistently observed right-skewed normal or lognormal path length distribution across all the studied cities.}
    \label{fig:supplementary_path_length_distribution}
\end{figure}

\begin{figure}[htp!]
    \centering
    \includegraphics[width=1\linewidth]{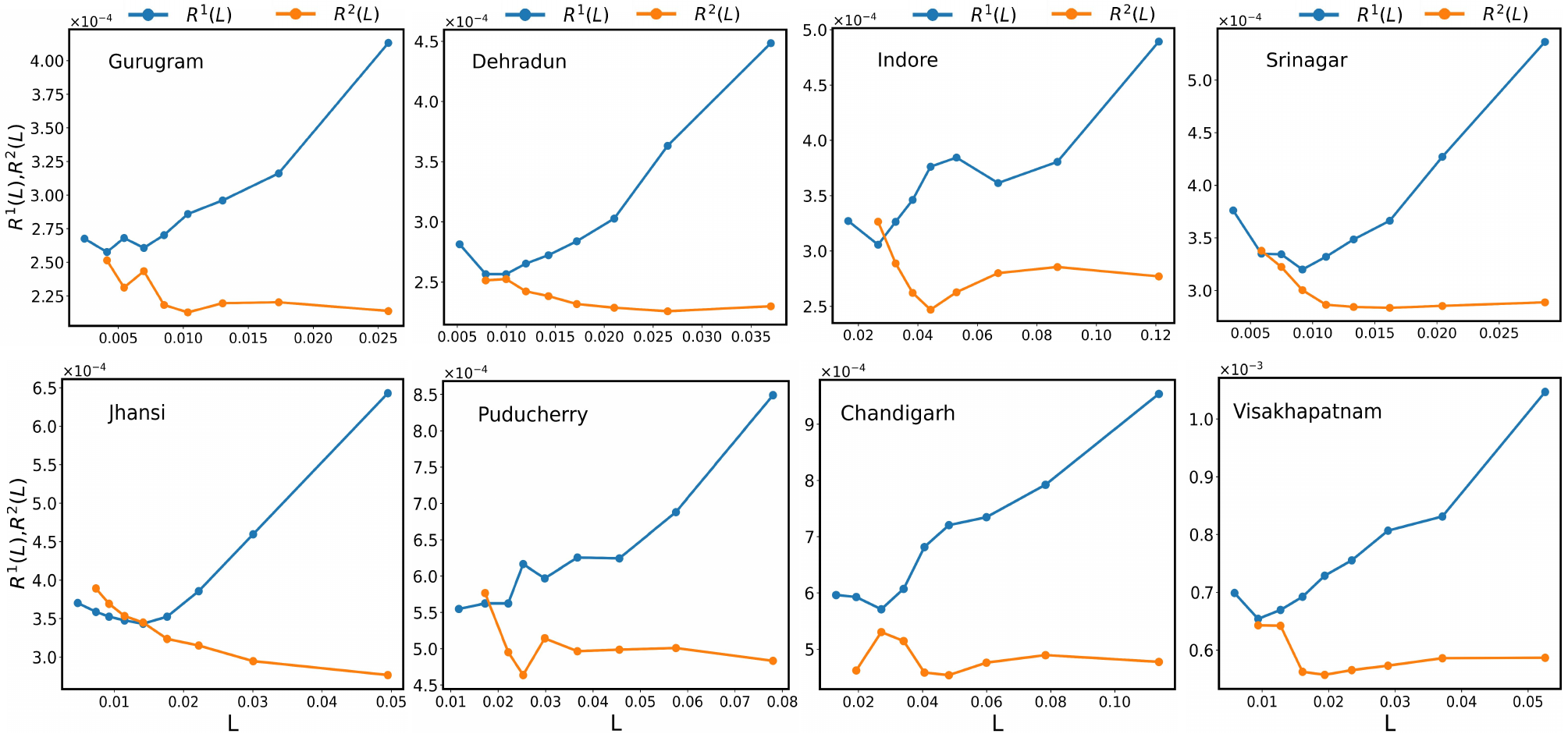}
    \caption{{\bf Rich Club of the long streets in SNs for the cities:} All the SNs follow the common trend of increase of $R^1$ and decrease of $R^2$ as discussed in the manuscript and indicate better connectivity of the shorter streets with the shorter and the longer streets with the similar or longer streets. }
    \label{fig:rich_club_supplemantary}
\end{figure}

As mentioned in section 2.2.1, the path lengths of the street networks for various cities exhibit right-skewed normal or lognormal behavior. Here, we present the path-length distribution of several other cities in Fig. \ref{fig:supplementary_path_length_distribution}.  Moreover, Fig.\ref{fig:rich_club_supplemantary} illustrates the rich club of long-range streets for a few more cities.

Further, table~\ref{tab:planarity_data} summarizes the planarity of the studied cities and the number of streets contributing to their non-planarity. 

\begin{table}[H]
    \centering
    \small
\begin{tabular}{|p{2.72cm}|p{2.72cm}|p{2.72cm}|p{2.72cm}|p{2.72cm}|}

 \hline
 City &  \#Nodes & \#Edges & Planar or Not & Number of Non-planar edges \\
 \hline
 Gurugram & 35580 & 47125 & False & 51\\
 \hline
 Dehradun & 33117 & 38295 & False & 77\\
 \hline
 Indore & 28998 & 40959 & False & 65\\
 \hline
 Gandhinagar & 27672 & 35839 & False & 64\\
 \hline
 Gwalior & 26916 & 35612 & False & 165\\
 \hline
 Srinagar & 26856 & 31687 & False & 101\\
 \hline
 Jhansi & 25552 & 31875 & False & 133\\
 \hline
 Jaipur & 25236 & 32276 & False & 27\\
 \hline
 Madurai & 23670 & 32106 & False & 27\\
 \hline
 Ambala & 23580 & 31359 & False & 68\\
 \hline
 Thanjavur & 20986 & 27778 & False & 85\\
 \hline
 Kollam & 20930 & 24474 & False & 78\\
 \hline
 Patna & 19034 & 24436 & False & 38\\
 \hline
 Kozhikode & 18204 & 22311 & False & 49\\
 \hline
 New Delhi & 17016 & 23055 & False & 34\\
 \hline
 Puducherry & 16683 & 22312 & False & 22\\
 \hline
 Chandigarh & 15426 & 21663 & False & 152\\
 \hline
 Vishakhapatnam & 14046 & 19662 & False & 47\\
 \hline
 Ujjain & 9131 & 12663 & False & 227\\
 \hline
 Imphal & 8691 & 10304 & True & \\
 \hline
 Navi Mumbai & 8364 & 11654 & False & 27\\
 \hline
 Ladakh & 8320 & 9962 & False & 78\\
 \hline
 Tirupati & 7642 & 10562 & False & 77\\
 \hline
 Aizwal & 6704 & 7749 & False & 21\\
 \hline
 Shimla & 6560 & 7018 & False & 30\\
 \hline
 Leh & 6246 & 7604 & False & 581\\
 \hline
 Guwahati & 5338 & 6705 & False & 21\\ 
 \hline
 SouthAndaman & 4144 & 4767 & True & \\
 \hline
 Kochi & 4051 & 4852 & False & 131\\
 \hline
 Kohima & 1782 & 2095 & True & \\
 \hline
 Daman & 1203 & 1506 & True & \\
 \hline
 Gangtok & 1130 & 1277 & False & 97\\
 \hline
 Dholera & 841 & 1183 & True & \\
 \hline

\end{tabular} 
\caption{{\bf Planarity of SNs:} This table summarizes whether the SNs of the city are planar or not. Also, it showed the number of streets making the SN non-planar. }
    \label{tab:planarity_data}
\end{table}

\section*{Spatial Arrangement of Street Networks}
\begin{figure}[htp!]
    \centering
    \includegraphics[width=1\linewidth]{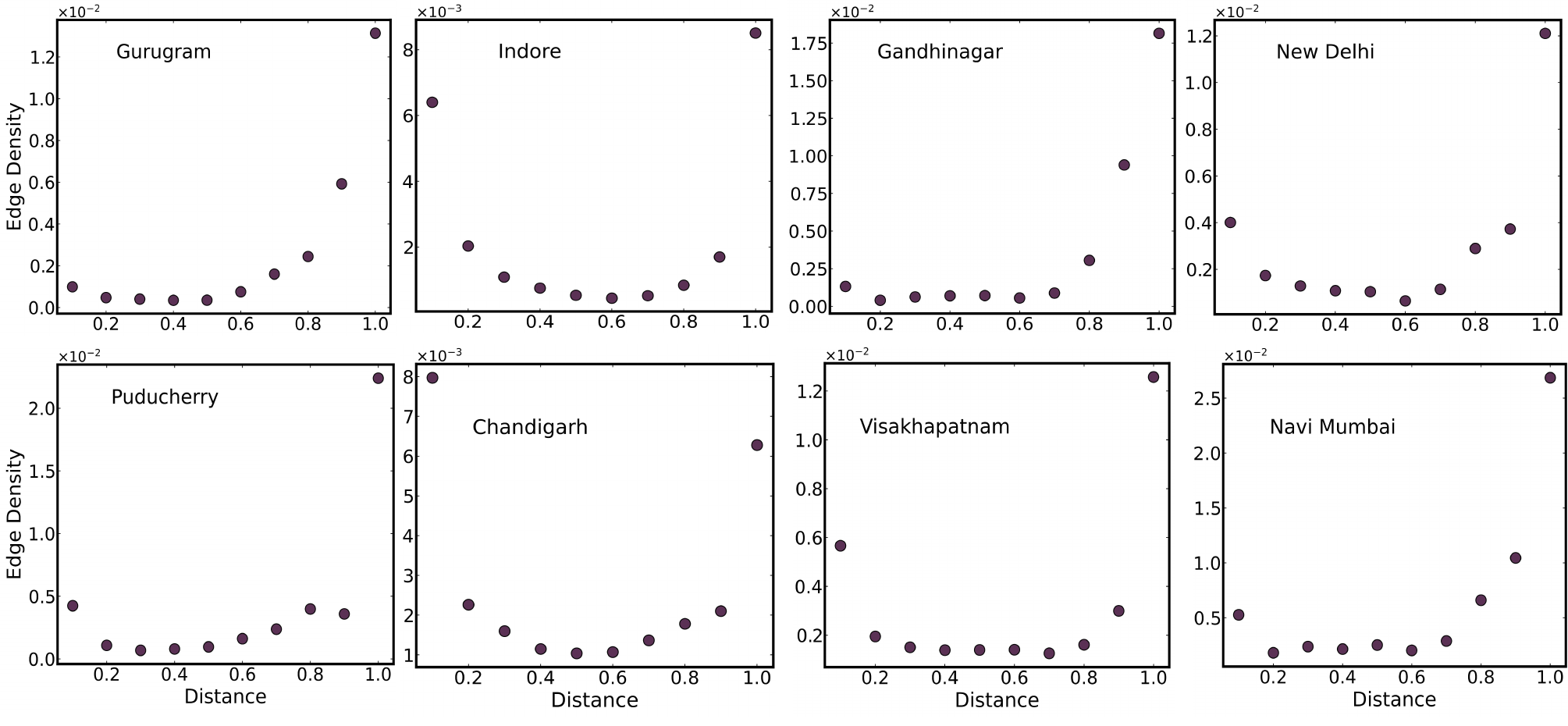}
    \caption{{\bf Variation of Edge Density from Barycenter:} The edge density increases as we move away from the barycenter, whereas, for some cities(mainly developed), the edge density is high near the barycenter and also near the periphery of the city.}
    \label{fig:edge_density_variation_barycenter}
\end{figure}

\begin{figure}[htp!]
    \centering
    \includegraphics[width=1\linewidth]{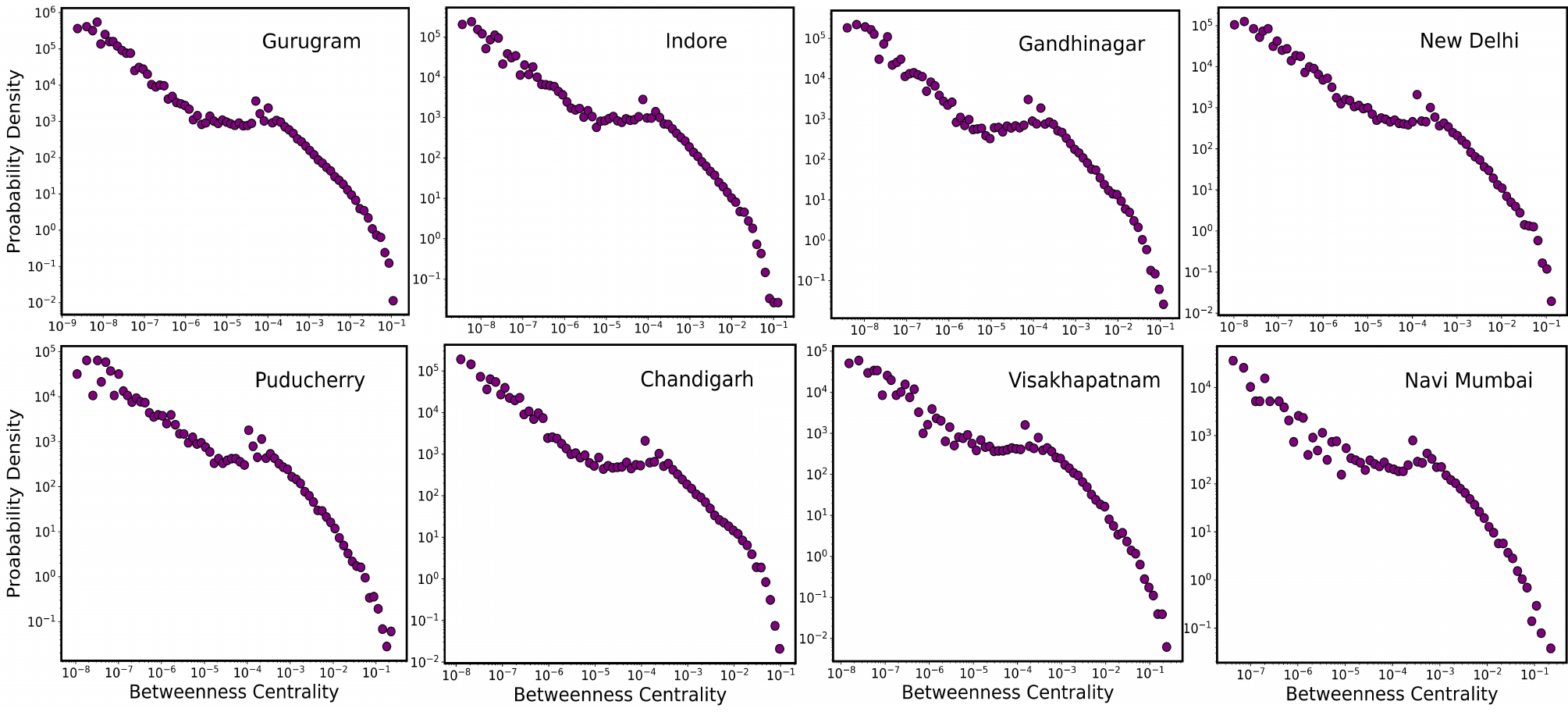}
    \caption{{\bf Betweenness Centrality Distribution:} The betweenness centrality distribution for different cities reveals the bimodal behavior of the street networks. We find similar behavior across all the Indian cities.}
    \label{fig:bc_distribution_supplementary}
\end{figure}

Orientation entropy, as discussed in the main manuscript, describes the global ordering of cities. A city's barycenter is calculated as the center of mass of a rigid body. To understand the spatial arrangement of SNs at the local level, we investigate the variation in betweenness centrality and edge density about the city's barycenter. We find that betweenness centrality is highest near the center and decreases towards the city's periphery(Fig. \ref{fig:bc_distribution_supplementary}). This pattern suggests that most crucial junctions are located near the barycenter of the cities, and travel between locations near the center often requires following specific routes, with limited or no alternative paths available. A similar result has been reported in \cite{from_bc_street_network}.
Furthermore, we find thatthe edge density generally increases towards the city's periphery, indicating strong inter-city connectivity with well-established links to neighboring towns or cities. However, there are some outliers. For some planned cities, we found that edge density is higher near the barycenter, decreases in the midsection, and then rises again at the city's periphery. This pattern suggests a well-connected street network within the city alongside robust connectivity to surrounding towns and cities, enhancing local and regional connectivity(Fig.\ref{fig:edge_density_variation_barycenter}).

\begin{figure}[htp!]
    \centering
    \includegraphics[width=1\linewidth]{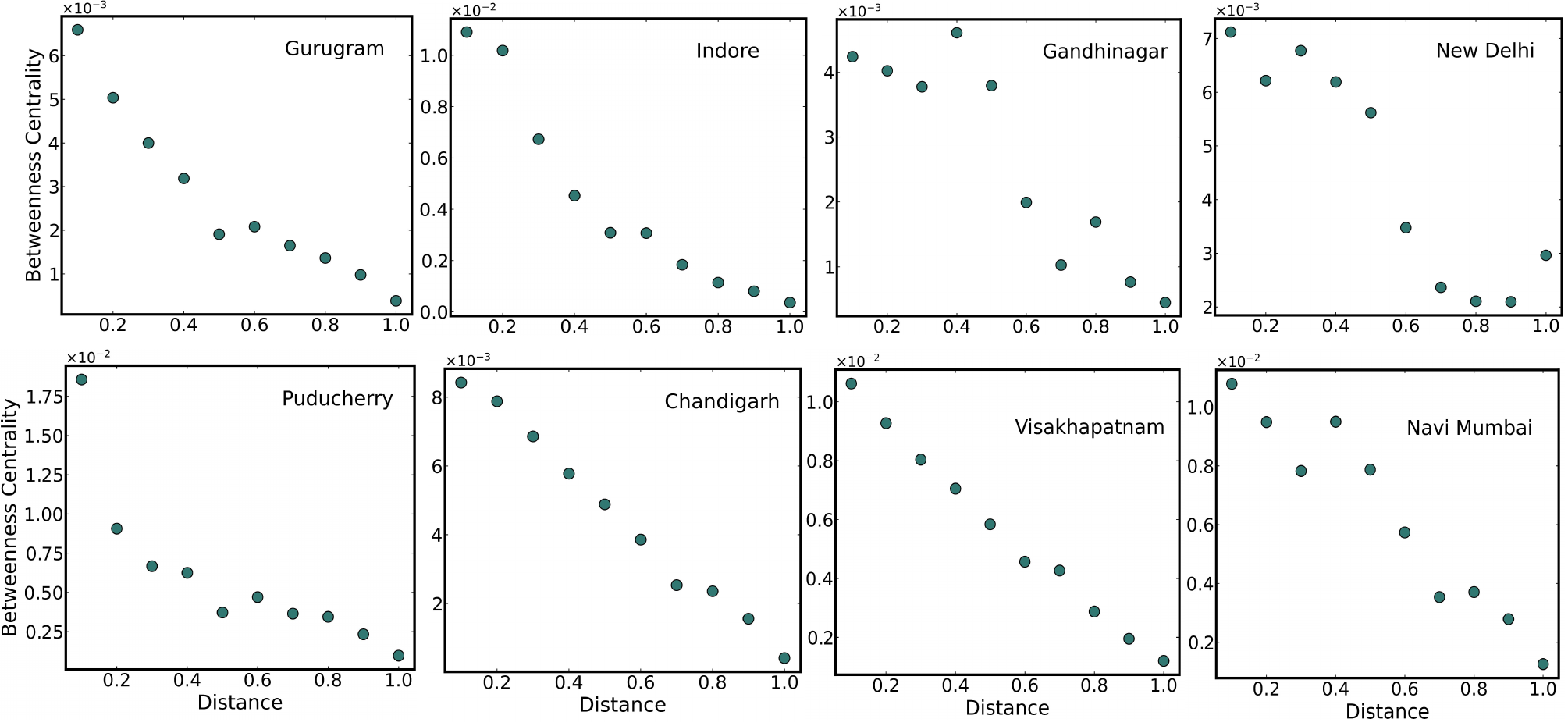}
    \caption{{\bf Variation of Betweenness centrality from Barycenter:} We observed a decrease in betweenness centrality as we moved away from the barycenter, indicating that the most important streets are located near the barycenter. This behavior remains consistent across all the cities that we studied.}
    \label{fig:bc_variation_barycenter}
\end{figure}

Further, we aim to check the betweenness centrality distribution of the studied cities. The betweenness centrality of the junction gives information about the number of the shortest paths passing through that particular junction. The betweenness centrality of the junction (node) $i$ can be given as :
\begin{equation}
    g_b(i) = \frac{2}{(N-1)(N-2)}\Sigma\frac{\sigma_{st(i)}}{\sigma_{st}},
    \nonumber
\end{equation}
where $\sigma_{st(i)}$ represents the edges between nodes $s$ and $t$ that are passing through $i$, and $\sigma_{st}$ are the total shortest path lengths between the nodes $s$ and $t$ \cite{networkx_betweenness_paper_algorithm,networkx_betweenness_first_paper}. We calculate the betweenness centrality of all the nodes for an SN and plot its distribution. Fig.\ref{fig:bc_distribution_supplementary} demonstrates the betweenness centrality distribution for some of the studied cities. We find that all the cities follow a bimodal behavior and fit well with the function: 
\begin{equation}
P(g_b) \sim {g_b}^{-\alpha}e^{-\frac{g_b}{\beta}}, 
\nonumber
\end{equation}
 where $g_b$ is the betweenness centrality as defined in the methods section, and $\alpha$ determines scaling behavior, whereas $\beta$ controls the cutoff of the distribution. This has been previously reported in another study \cite{from_bc_street_network}. 

\begin{figure}[htp!]
    \centering
    \includegraphics[width=0.62\linewidth]{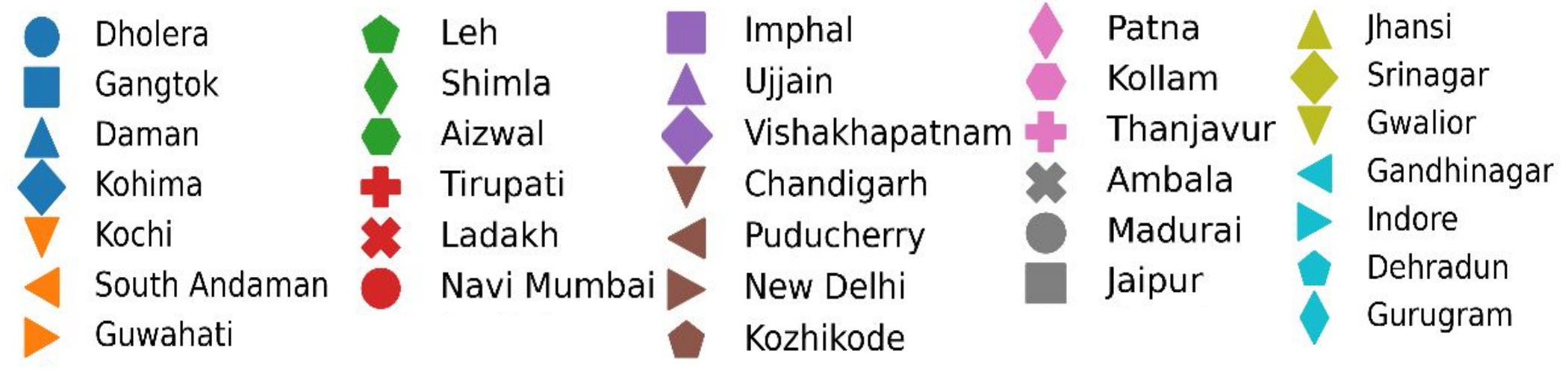}
    \caption{ {\bf Symbols for cities:} Different markers and colors represent different cities, each with a unique combination of markers and colors. }
    \label{fig:city_labels}
\end{figure}

Furthermore, as shown in Fig. \ref{fig:city_labels}, we have included a comprehensive representation  f the markers and color codes assigned to each city. These specific markers and colors align with the data presented in both Fig. \ref{fig:Small_world_behavior} and Fig. \ref{fig:rcc_empirical}.

\end{document}